\documentstyle[11pt-{article}

\newlength{\dinwidth}
\newlength{\dinmargin}
\def\st{\widetilde{t}}

\def\stl{\st_{1}}

\def\mstl{m_{\stl}}
\def\sth{\st_{2}}

\def\mz{m_{Z}}

\def\tht{\theta_{t}}
\def\tew{\theta_{W}}
\def\slepton{\widetilde \ell}

\def\msl{m_{\slepton}}

\def\wino{{\widetilde{\rm W}}}
\def\sfermion{\widetilde f}
\def\sf{\widetilde f}
\def\sfb{\widetilde{\overline f}}

\def\sele{\widetilde e}
\def\ser{\sele_{R}}

\def\smu{\widetilde \mu}
\def\stau{\widetilde \tau}
\def\staul{\stau_{1}}
\def\su{\widetilde{u}}
\def\sd{\widetilde{d}}

\def\ddf{{\rm d}}
\def\bino{\widetilde{B}}
\def\higgsino{\widetilde{H}^{0}}
\def\sz1{{\widetilde{Z}}_{1}}
\def\swl{{\widetilde{W}}_{1}}

\def\mser{m_{\ser}}
\def\msz1{m_{\sz1}}
\def\mswl{m_{\swl}}
\def\mbino{m_{\bino}}
\def\gev{{\rm GeV}}

\def\rs{{\sqrt{s}}}
\def\Ee{E_{e}}
\def\Eet{E_{eT}}
\def\te{\theta_{e}}

\def\nle{{\stackrel{<}{\sim}}}
\def\nge{{\stackrel{>}{\sim}}}
\def\goto{\rightarrow}
\def\sh{{\hat{s}}}
\def\bh{{\hat{\beta}}_{f}}
\def\rsh{\sqrt{\sh}}

\begin{document}
{}~~~\\
\begin{flushright}
ITP-SU-93/02 \\
\end{flushright}
\begin{center}
  \begin{Large}
  \begin{bf}
\renewcommand{\thefootnote}{\fnsymbol{footnote}}
SUSY at High Energy $e\gamma$ and $\gamma\gamma$ Colliders
\footnote{
      Based on talks presented at
{\it the 4th Workshop on JLC}, (KEK, 15--16, March, 1993)
and
{\it the 2nd International Workshop on Physics
and Experiments at Linear $e^{+}e^{-}$ Colliders},
(Waikoloa, Hawaii, 26-30, April, 1993)
}
 \\
  \end{bf}
  \end{Large}
  \vspace{5mm}
  \begin{large}
\renewcommand{\thefootnote}{\fnsymbol{footnote}}
    Tadashi Kon
\footnote{
e-mail address : d34477@jpnac.bitnet, mtsk@jpnkektr.bitnet
}
\\
  \end{large}
Faculty of Engineering, Seikei University, Tokyo 180, Japan
  \vspace{5mm}
\end{center}
\begin{quotation}
\noindent
{\bf Abstract:}
We investigate the sparticle production processes
$e\gamma\goto \sele\sz1$ and
$\gamma\gamma\goto \sfermion\overline{\sfermion}$
at high energy $e\gamma$ and $\gamma\gamma$ colliders
in the framework of the minimal supersymmetric standard model
(MSSM).
It will be shown that the $e\gamma$ colliders would be more suitable
in searching for the heavy selectrons than $ee$ colliders because
of the low mass threshold of the process
$e\gamma\goto \sele\sz1$.
We show that the standard background processes
$e\gamma\goto \nu W$ and $eZ$ can be suppressed in terms of
initial beam polarization as well as the kinematical
cuts on the energy and angle of the final electron.
Moreover, it will be argued that
the experimental measurements
of the cross sections
for the processes
$e\gamma\goto \sele\sz1$ and
$\gamma\gamma\goto \sfb\sf$
could enable us to constrain the basic parameters
in the MSSM.
This is originated from the simplicity of SUSY parameter dependence
of these processes.
We also give comments on the stop production at
$e^{+}e^{-}$ colliders
with the longitudinally polarized electron beams.

\end{quotation}
\section{Introduction}
In addition to well-known types of "TeV" colliders such as
$pp$, $ep$ and linear $e^{+}e^{-}$ colliders,
the possibilities for realization of the
$e\gamma$ and $\gamma\gamma$ colliders have been discussed in
detail by Ginzburg et al. \cite{Ginz}.
Here the high energy photon beams will be obtained by the
backward Compton scattering of the laser flush by one of
electron beam in the basic linear $ee$ colliders.
Such colliders will provide us with a powerful
machinery for investigating the standard model through a direct
proof of the gauge vertices \cite{YH} and
searching for the neutral Higgs boson \cite{Higgsgg,Higgseg,HWZ}.
The $e\gamma$ and $\gamma\gamma$ colliders are
also suited for searching for
some exotic particles predicted by the models beyond the standard model,
such as supersymmetric partners \cite{ours,susy}
and excited fermions \cite{GI,egest}.
In this talk I will present recent results of the
analyses for the SUSY particle (sparticle) production at
$e\gamma$ and $\gamma\gamma$ colliders,
which are based on the previous our works \cite{ours}.

\section{Calculation}

We calculate the experimental cross sections folding the sub-process
cross section
$\hat{\sigma}$($e\gamma\goto X$) with
the energy spectra $D_{\gamma/e}$
of the high energy photons generated by
the backward Compton scattering of the laser light ;
\begin{equation}
\sigma(e\gamma\goto X)=
\int\ddf zD_{\gamma/e}(z)
{\hat{\sigma}}(e\gamma\goto X ; \xi_{2}(z)),
\end{equation}
where $D_{\gamma/e}$ and $\xi_{2}$ denote respectively
the photon energy spectrum
and the mean helicity (Stokes parameter)
of the high energy photons \cite{Ginz}.
Here $z$ denotes the energy fraction of the high energy photons,
in other words, the ratio of total energies squared in
the $e\gamma$ and $ee$ collisions ;
$z=\sh/s\equiv{s_{e\gamma}}/{s_{ee}}$.
Note that the upper limit of $z$ is determined by the kinematics
of the backward-Compton scattering, i.e., $z\nle 0.83$.
If $z$ becomes larger than this value the back-scattered and
laser photon have enough energy to produce the $e^{+}e^{-}$ pair,
and in turn the conversion efficiency drops considerably \cite{Ginz}.
In the following we take $z_{max}=0.83$.
For the $\gamma\gamma$ collision which will be discussed in Sec~4,
the experimental cross sections can be obtained by
\begin{equation}
\sigma(\gamma\gamma\goto X)=
\int\ddf z_{1}\int\ddf z_{2}
D_{\gamma/e}(z_{1})D_{\gamma/e}(z_{2})
{\hat{\sigma}}(\gamma\gamma\goto X).
\end{equation}
The maximum value of the total energies of $e\gamma$ and
$\gamma\gamma$ sub-processes correspond to about
90\% and 80\% of $\rs$ of the basic $ee$ colliders,
\begin{eqnarray}
\sqrt{s_{e\gamma}}&<&\sqrt{z_{max}s_{ee}}=
\sqrt{0.83s_{ee}}\simeq0.9\rs \\
\sqrt{s_{\gamma\gamma}}&<&\sqrt{{z_{1}}_{max}{z_{2}}_{max}s_{ee}}\simeq
0.8\rs.
\label{zmax}
\end{eqnarray}

\section{Selectron Production in $e\gamma$ Collisions}

First I'll discuss the single selectron production in the
electron-photon colliders.
Here we pay attention to the right-handed selectron $\ser$ production,
because this mass eigenstate is lighter than the left-handed one
in most of the supergravity GUTs.
It is expected, moreover,
that the signature of this process will be rather simple
because $\ser$ decays dominantly into
$e \sz1$ \cite{tsuka}, where $\sz1$ denotes the lightest neutralino.
Since we can assume $\sz1$ as the LSP, we find that
this process has lower mass threshold $\mser+\msz1$ than
that $2 \mser$ of the selectron pair production at
$e^{+}e^{-}$ colliders.
Consequently, it is expected that the $e\gamma$ colliders
will be efficient in searching for the heavy selectron with mass
larger than half of $\rs$ in $e^{+}e^{-}$ colliders.
Figure 1 show the selectron mass dependence of the total
cross sections.
\begin{figure}[t- \label{FIGUR1}
  \vspace{7cm}
  \caption[strunt-{{
Slepton mass dependence of total cross sections.
We take $\rs=300\gev$ and $\msz1=\mbino=50\gev$.
  }}
\end{figure}
We can obtain the large cross section for
$e\gamma$ $\goto$ $\ser\sz1$ even if the selectron is
rather heavy $\mser\nge \rs /2$.
Note that the initial beam polarization will be efficient
to enhance the signal cross section.

Another important property of this process is the simple SUSY
parameter dependence of the cross section.
The cross section is proportional to the bino component $N_{11}$
of the lightest neutralino
$\sz1 = N_{11}\bino+N_{12}\wino+N_{13}\higgsino_{1}+N_{14}\higgsino_{2}$.
The arbitrary SUSY parameters appeared in the cross section
are $\mser$, $\msz1$ and $!N_{11}!$ only.
This is in contrast to the rather complicate situation
in the selectron pair production at $e^{+}e^{-}$ colliders,
to which all masses and many mixing angles of the neutralinos
contribute.
If we know $\mser$ and $\msz1$ from the analysis on the process
$e^{+}e^{-}$ $\goto$ $\ser^{*}\ser$ \cite{tsuka}
we can determine $!N_{11}!$ by the measurement of total cross section
for $e\gamma$ $\goto$ $\ser\sz1$.
Since $!N_{11}!$ has unique contour in the
($\mu$, $M_{2}$) plane depicted in Fig.~2,
\begin{figure}[t- \label{FIGUR2}
  \vspace{7cm}
  \caption[strunt-{{
Contours of $!N_{11}!$ and $\msz1$
in ($\mu$, $M_{2}$) plane. We take
$\tan\beta$ $=$ 10.
  }}
\end{figure}
we will get valuable information on the SUSY parameters
$\mu$ and $M_{2}$.

Next we should discuss the background suppression.
Since the signature of our process will be the single electron
plus large missing energies taken away by the LSP,
two types of the SM processes,
$e\gamma\goto W\nu \goto (e{\overline{\nu}})\nu$ and
$e\gamma\goto eZ \goto e({\overline{\nu}}\nu)$,
will become serious backgrounds,
as we can see in Fig.~3.
\begin{figure}[t- \label{FIGUR3}
  \vspace{7cm}
  \caption[strunt-{{
$\sqrt{s_{e\gamma}}$ ($\rsh$) dependence of
total cross sections.
We take $\mser=200\gev$ and $\msz1=\mbino=50\gev$ for
$e\gamma$ $\goto$ $\ser\sz1$.
  }}
\end{figure}
I will show that these backgrounds can be suppressed
by 1) initial beam polarization and 2) appropriate kinematical
cuts on the energies and angles of scattering electron.
In fact, we find in Fig.~4 that the $W\nu$ process can be suppressed
if we take the right-handed initial electron beams
($P_{e_{R}}$=$P_{e'_{R}}$=0.95)
and the minus helicity laser beam ($P_{c}$=-1),
where we set the degree of
longitudinal polarization of electrons as 95\%.
\begin{figure}[t- \label{FIGUR4}
  \vspace{7cm}
  \caption[strunt-{{
Scattering angle distribution of differential cross section.
We take $\rs=300\gev$, $\mser=200\gev$ and $\msz1=\mbino=50\gev$.
Beam polarization are taken as
[$P_{e_{R}}$, ($P_{e'_{R}}$, $P_{c}$)- $=$ [$0.95$,($0.95$, $-1$)-.
  }}
\end{figure}
Moreover, the appropriate cuts on the energy $\Ee$,
the transverse energy $\Eet$ and the energy fraction $z$ can be
use to suppress the $eZ$ process (see Fig.~5).
\begin{figure}[t- \label{FIGUR5}
  \vspace{7cm}
  \caption[strunt-{{
Scattering angle distribution of differential cross section
(Same setup with Fig.~4) after kinematical cuts,
$\Ee < 100\gev$, $\Eet > 60\gev$ and $z > 0.7$.
  }}
\end{figure}
Note that $z$ can be reconstructed
from $\Ee$ and $\cos\te$.
After taking such kinematical cuts, we will find the excess of the
cross section at the backward direction as the SUSY signal.

\section{Sfermion Production in $\gamma\gamma$ Collisions}

Next topic is the sfermion pair production at the
$\gamma\gamma$ colliders.
Since these are pure SUSY QED processes,
we can describe cross sections for all sfermions
$\sfermion =$ ($\sele_{L,R}$, $\smu_{L,R}$, $\stau_{1,2}$;
$\su_{L,R}$, $\sd_{L,R}$, $\cdots$, $\st_{1,2}$)
by the formula :
\begin{equation}
{\hat{\sigma}}(\gamma\gamma\goto\sfb\sf)=
{\frac{2\pi\alpha^2}{\sh}}{e_{f}^{4}}C_{f}\bh
\left(2-\bh^{2}-{\frac{1-\bh^{4}}{2\bh}}\ln{\frac{1+\bh}{1-\bh}}\right),
\end{equation}
where
$e_{f}$ and $C_{f}$ respectively denote the electric charge
and color degree of freedom and
$\bh=\sqrt{1-4m_{\sfermion}^{2}/\sh}$.
One of good properties of these processes is the simple
dependence on the arbitrary SUSY parameters, i.e.,
the cross sections depend only on the final sfermion masses.
This could be useful to check the universality of masses
of sleptons and squarks in the 1st and 2nd generation.

While the mass threshold for these processes will be
larger than that of $e^{+}e^{-}\goto\sfb\sf$
because of the upper limit Eq.(\ref{zmax})
of the photon energy fraction $z$,
the large cross section compared to $e^{+}e^{-}$ case
is expected owing to the $s$-wave production,
in other words, the typical $\bh$ dependence.
It should be noted that this property also enables us to see the
squarkonium production.
The slepton mass dependence of total cross section is
shown in Fig~6.
\begin{figure}[t- \label{FIGUR6}
  \vspace{7cm}
  \caption[strunt-{{
Slepton mass dependence of total cross section.
We take $\rs=500\gev$.
  }}
\end{figure}
The selectron production at $e^{+}e^{-}$ colliders has large
cross section originated from existence of the $t$-channel neutralino
exchange diagram.
On the other hand, the $\gamma\gamma$ cross section will be much
larger than $e^{+}e^{-}$ one for 2nd and 3rd generation sleptons
if $\rs$ is large enough compared with the mass threshold,
$\rs$$\nge$$3m_{\sfermion}$.
This is also true for the stop production as shown in Fig.~7,
where I have included the QCD radiative correction \cite{bigi}.
\begin{figure}[t- \label{FIGUR7}
  \vspace{7cm}
  \caption[strunt-{{
$\rs$ dependence of total cross section for
$\mstl=120\gev$ and $200\gev$.
  }}
\end{figure}

Here I will give comments on the scalar top (stop) production at
ordinary $e^{+}e^{-}$ colliders \cite{DH}.
One of the stop $\stl$ will be lighter than
the other squarks and possibly lighter than the top itself
because of the large top Yukawa coupling and the left-right
mixing of the stops \cite{stop}.
The mass eigenstates are parametrized by the mixing angle $\tht$ ;
\begin{equation}
\left({\stl\atop\sth}\right)=
\left(
{\st_{L}\,\cos\tht-\st_{R}\,\sin\tht}
\atop
{\st_{L}\,\sin\tht+\st_{R}\,\cos\tht}
\right)
\end{equation}
and their mass eigenvalues can be expressed by
\begin{equation}
m^{2}_{\stl\atop\sth}
         ={\frac{1}{2}}\left[ m^{2}_{\st_{L}}+m^{2}_{\st_{R}}
             \mp \left( (m^{2}_{\st_{L}}-m^{2}_{\st_{R}})^{2}
            +(2a_{t}m_{t})^{2}\right)^{1/2}\right-,
\label{stopmass}
\end{equation}
where $m_{\st_{L, R}}$ and $a_{t}$ are the SUSY mass parameters.
We find that if SUSY mass parameters and the top
mass are the same order of magnitude, cancellation could
occur in the expression for the
lighter stop mass Eq.~(\ref{stopmass}).
So we could get one light stop $\stl$ lighter than the top as well as
the other squarks.
It has been shown that such light stop with mass
$\mstl$ $<$ $\mswl$, $\msl$ has dominant decay mode
$\stl \goto c \sz1$ \cite{stop}.
We note that
the stop with mass $\mstl$$\nle$$30\gev$
has not still been excluded
if the mixing angle $\tht$ is in the range
$0.8$$\nle$$\tht$$\nle$$1.2$ and if $m_{\sz1}$$\nge$$10\gev$
even considering negative search at TEVATRON \cite{Baer}
and LEP \cite{DH,Fisher}.
Consequently, there remains oppotunity to discover such light stop
even for TRISTAN \cite{stop,TRISTAN} and/or HERA \cite{HERA}.
We note that the existence of such light stop will
alter the signatures of the top quark \cite{Baer,HKK}
and the Higgs bosons \cite{AHKKM}.

Here I show that the stop mixing angle $\tht$
as well as the stop mass $\mstl$ can be
measured at the linear $e^{+}e^{-}$ colliders,
if we use the longitudinally polarized electron beams.
The total cross sections with the longitudinally polarized beams
can be expressed by
\begin{equation}
\sigma(e_{L\atop R}) =
{\frac{\pi\alpha^{2}}{s}}\beta_{\stl}^{3}
\left[{\frac{4}{9}}
+C_{\stl}^{2}A_{L\atop R}^{2}
\left!{\frac{s}{D_{Z}}}\right!^{2}
+{\frac{4}{3}}C_{\stl}A_{L\atop R}{\rm Re}
\left({\frac{s}{D_{Z}}}\right)\right-(1+\delta_{QCD}),
\end {equation}
where
$\beta_{\stl}=\sqrt{1-4\mstl^{2}/s}$,
$D_{Z}\equiv s-\mz^2+i\mz\Gamma_{Z}$,
$A_{L}\equiv (1-2\sin^{2}\tew)/\sin^{2}2\tew$,
$A_{R}\equiv (-2\sin^{2}\tew)/\sin^{2}2\tew$
and $C_{\stl}\equiv\cos^{2}\tht-{\frac{4}{3}}\sin^{2}\tew$.
{}From the formula we can find that the left-right asymmetry
\begin{equation}
A_{LR}\equiv{\frac{\sigma(e_{L})-\sigma(e_{R})}
                  {\sigma(e_{L})+\sigma(e_{R})}}
\end{equation}
depends sensitively on $C_{\stl}$ and in turn on $\tht$,
because $A_{LR}$ is proportional to $C_{\stl}$.
It should be noted that $A_{LR}$ will be independent
on the mass of stop ($\beta_{\stl}$) as well as
the QCD radiative correction $\delta_{QCD}$.
In Fig~8 we show the $\tht$ dependence of the total cross section
and the left-right asymmetry.
\begin{figure}[t- \label{FIGUR8}
  \vspace{7cm}
  \caption[strunt-{{
Stop mixing angle dependence of total cross section and
left-right asymmetry for $e^{+}e^{-}$ $\goto$ $\stl^{*}\stl$.
Scattering angle distribution of differential cross section.
We take $\rs=300\gev$.
  }}
\end{figure}
The total cross section depends sensitively on the mass but
not on $\tht$ at $\rs$ larger than a few handred GeV.
On the other hand, the dependence of $A_{LR}$ on $\tht$ is very
sensitive.
So we can conclude that the total cross section and the
left-right asymmetry will be complementally useful to measure
the mass and the mixing angle of the stop.
We expect that the mixing angle measured experimentally will
bring us valuable information of the basic
SUSY parameters such as the trilinear coupling constant $A_{t}$.
Note that we will also know the mixing angle $\theta_{\tau}$
of the stau $\stau$ by measuring $A_{LR}$ for
$e^{+}e^{-}$ $\goto$ $\staul^{*}\staul$.

As mentioned above, we may see the stoponium \cite{stoponium}
resonance peak in the invariant mass distribution
at $\gamma\gamma\goto \stl^{*}\stl$
shown in Fig.~9.
\begin{figure}[t- \label{FIGUR9}
  \vspace{7cm}
  \caption[strunt-{{
Invariant mass distribution of differential cross section.
We take $\rs=500\gev$ and $\mstl=200\gev$.
  }}
\end{figure}
{}From the peak structure we may extract valuable information
on the SUSY parameters as well as on the SUSY QCD
because the total width of stoponium depends sensitively on
the arbitrary SUSY parameters.

\section{Conclusion}
\label{sec:conclusion}

  I have investigated the sparticle production in $e\gamma$ and
$\gamma\gamma$ colliders.
The single selectron production has lower mass threshold than
the pair production at $e^{+}e^{-}$ or $\gamma\gamma$ colliders
and will be suitable in searching for the heavy selectrons.
I have shown that the SM backgrounds $W\nu$ and $eZ$ could be
suppressed by initial beam polarization as well as appropriate
kinematical cuts.
In addition, the process has rather simple SUSY parameter dependence
and in turn we could know the $\bino$ component of the lightest
neutralino in terms of the cross section experimentally measured.
We have also given comments on the stop production at
$e^{+}e^{-}$ colliders.
The stop mixing angle $\tht$
as well as the stop mass $\mstl$ can be
measured at the linear $e^{+}e^{-}$ colliders
if we use the longitudinally polarized electron beams.
The longitudinally polarized asymmetry $A_{LR}$ depends sensitively
on $\tht$.
We expect that the mixing angle measured experimentally will
bring us valuable information of the basic SUSY parameters.
In the $\gamma\gamma$ colliders the cross section can be larger than
$e^{+}e^{-}$ one if $\rs$ is large enough, $\rs\nge 3m_{\sfermion}$,
for $\smu$ and $\stau$.
The same origin, the $s$-wave production of the sfermion pair,
could enable us to see the unique squarkonium
(especially the stoponium) resonance.
It has also been commented that we may use the cross sections
in order to check the universality of the sfermion masses.
Based on above discussion, my conclusion is
$e\gamma$ and $\gamma\gamma$ options will be efficient in
searching for sparticles and also in determining SUSY parameters
in the MSSM.

\vskip 20pt
\begin{flushleft}
{\Large{\bf Acknowledgement}}
\end{flushleft}
The author thanks A. Goto for his collaboration at early stage
of this work.

\end{document}